# Rate Analysis or a Possible Interpretation of Abundances

**Miklós Kiss[1]**

*Berze High School/Gyöngyösi Berze Nagy János Gimnázium*
*H-3200 Gyöngyös, Kossuth u. 33., Hungary*
*E-mail:* `kiss-m@chello.hu`

Heavy elements are formed in nucleosynthesis processes. Abundances of these elements can be classified as elemental abundance, isotopic abundance, and abundance of nuclei. In this work the nuclei are identified by (Z,N), which allows reading out *new information* from the measured abundances. We are interested in the neutron density required to reproduce the measured abundance of nuclei assuming equilibrium processes. This is only possible when two stable nuclei are separated by an unstable nucleus. At these places we investigated the neutron density required for equilibrium nucleosynthesis both isotopically and isotonically at temperatures of AGB interpulse and thermal pulse phases. We obtained an estimate for equilibrium nucleosynthesis neutron density in most of the cases. Next we investigated the possibility of partial formation of nuclei. We analyzed the meaning of the branching factor. We found a mathematical definition for the unified interpretation of a branching point closed at isotonic case and open at isotopic case. We introduce a more expressive variant of branching ratio called partial formation rate. With these we are capable of determining the characteristic neutron density values. We found that all experienced isotope ratios can be obtained both at $10^8 \, \mathrm{K}$ temperature and at $3 \cdot 10^8 \, \mathrm{K}$ temperature and at intermediate neutron density ($\leq 2 \cdot 10^{12} \, \mathrm{cm}^{-3}$).

---

[1]

Speaker

## 1. Introduction

Nearly sixty years after BBFH [1], it is possible and necessary to review and rethink our knowledge about the neutron capture nucleosynthesis. The result of the formation of the nuclei is shown in the abundances. It is important to mention that the formed unstable nuclei have decayed into stable nuclei and we are only able to see the resulting stable nuclei.

"The success of any theory of nucleosynthesis has to be measured by comparison with the abundance patterns observed in nature." say Käppeler, Beer and Wisshak [2], that is, we need to create such model that gives back the observed abundance.

Because of the formation of nuclei takes place in a variety of conditions, the experienced abundance is a result of more processes. Therefore more models are necessary for the alternate conditions. According to the conditions of the models the nuclei are classified into categories as s-nuclei, r-nuclei etc.

It seems that the reverse approach is also useful: the abundance is the preserver of the nuclei's formation conditions. So instead investigating whether the theoretical model fits the observed abundance, we look for the circumstances when the observed abundance is available.

To do this we need suitable data: the half-life of unstable nuclei and the neutron capture cross section of nuclei. These data are not constant always. At some nuclei the half-lives depend on the temperature [2,3,4]. Fortunately, the reaction rate per particle pair $< \sigma v >$ is constant between 10 and 100 keV because of the energy dependence of $\sigma$ [2,3]. So we can use the $\sigma$ values at 30 keV [5]. The possible resonances only improve the capture capabilities.

## 2. The required neutron density

Change the nucleon identification from the usual (Z,A) to (Z,N) and see the individual abundances as well. We took the abundance of nuclei from [6]. This will allow us to read new information from the various measured abundances.

We use the following rate equations according to the requirements of the individual nuclei formation:

$$\frac{dN_{Z,N}}{dt} = N_n(t)N_{Z,N-1}(t) < \sigma v >_{Z,N-1} + \lambda_\beta N_{Z-1,N+1}(t) + \lambda_\alpha N_{Z+2,N+2}(t) - $$
$$- N_n(t)N_{Z,N}(t) < \sigma v >_{Z,N} - \lambda_\beta N_{Z,N}(t) - \lambda_\alpha N_{Z,N}(t) + \ldots$$

We also assume the equilibrium formation of nuclei. From the corresponding rate equations we can get the neutron density as in isotopic as in isotonic cases.

### 2.1 Isotopic case

#### 2.1.1 Two stable neighbor isotopes

First let us we consider two stable neighbor isotopes (see Fig. 1), applying the new identification.

$$\frac{dN_2}{dt} = n_n N_1 < \sigma v >_1 - n_n N_2 < \sigma v >_2$$

$$\frac{dN_2}{dt} = n_n v(N_1 \sigma_1 - N_2 \sigma_2)$$

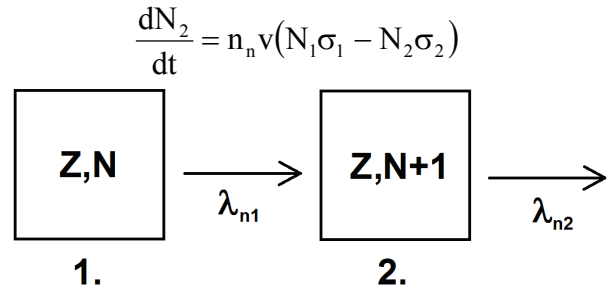

*Fig. 1. Two stable neighbor isotopes*

If we suppose the equilibrium that is $\frac{dN_2}{dt} = 0$, we got $N_1 \sigma_1 = N_2 \sigma_2$, what is well known in classical identification. So there is no information about the formation conditions. But it is important to mention that this relation is not true in general about the experienced abundances and neutron capture cross sections. The main cause is the existences of other rates from other channels through unstable nuclei.

### 2.1.2 Two stable isotopes are separated by an unstable isotope

A more interesting case is, when two stable isotopes are separated by an unstable isotope (see Fig. 2).

In this isotopic case the rate equations for the unstable nucleus:

$$\frac{dN_2}{dt} = n_n N_1 <\sigma v>_1 - n_n N_2 <\sigma v>_2 - \lambda_\beta N_2$$

About the equilibrium $\frac{dN_2}{dt} = 0$, so

$$0 = n_n (N_1 <\sigma v>_1 - N_2 <\sigma v>_2) - \lambda_\beta N_2$$

$$\lambda_\beta N_2 = n_n (N_1 <\sigma v>_1 - N_2 <\sigma v>_2)$$

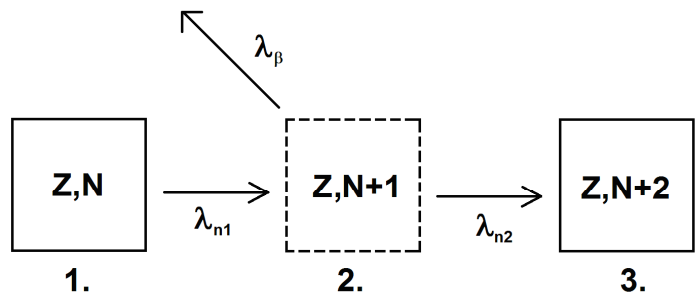

*Fig. 2. The isotopic channel*

$$\lambda_\beta N_2 = n_n v(N_1 \sigma_1 - N_2 \sigma_2)$$

$$\lambda_\beta = n_n v \left( \frac{N_1}{N_2} \sigma_1 - \sigma_2 \right)$$

If the third nucleus can form only from the second nucleus

$$\frac{dN_3}{dt} = n_n N_2 < \sigma v >_2 - n_n N_3 < \sigma v >_3$$

About the equilibrium $\frac{dN_3}{dt} = 0$, so such as the previous case we have

$$N_2 \sigma_2 = N_3 \sigma_3$$

$$N_2 = N_3 \frac{\sigma_3}{\sigma_2}$$

$$\lambda_\beta = n_n v \left( \sigma_1 \frac{N_1 \sigma_2}{N_3 \sigma_3} - \sigma_2 \right)$$

$$\frac{\lambda_\beta}{n_n v} + \sigma_2 = \frac{N_1 \sigma_1 \sigma_2}{N_3 \sigma_3}$$

$$\frac{N_3}{N_1} = \frac{\sigma_1 \sigma_2}{\left( \frac{\lambda_\beta}{n_n v} + \sigma_2 \right) \sigma_3}$$

$$\frac{N_3}{N_1} = \frac{< \sigma v >_1}{< \sigma v >_3} \frac{< \sigma v >_2}{\frac{\lambda_\beta}{n_n} + < \sigma v >_2} = \frac{\sigma_1}{\sigma_3} \frac{\sigma_2}{\frac{\lambda_\beta}{n_n v} + \sigma_2}$$

This expression contains the specific neutron density value. We must refer to
$< \sigma v > = \sigma \cdot v = \text{constant}$ at low and intermediat conditions, so it is enough to know the
30 keV neutron capture cross sections and average velocity.
From this formula, we can get the neutron density value. Of coure resonances may occure, but
these only improve the capture capabilities

$$n_n = \frac{\lambda_\beta}{v \left( \sigma_1 \frac{N_1 \sigma_2}{N_3 \sigma_3} - \sigma_2 \right)} \text{ or } n_n = \frac{\lambda_\beta}{v} \frac{\sigma_3}{\sigma_2} \frac{1}{\left( \sigma_1 \frac{N_1}{N_3} - \sigma_3 \right)}$$

(See the required neutron densities at different stellar temperature in Appendix 1 and 2.)
From the tables we can see that the isotopic equilibrium is not always possible. This means that
the formation of third nuclei has at least another channel.

**2.2 Isotonic case**

In this isotopic case (see Fig. 3) the rate equations for the unstable nucleus:

$$\frac{dN_2}{dt} = n_n N_1 < \sigma v >_1 - n_n N_2 < \sigma v >_2 - \lambda_\beta N_2$$

If there is equilibrium, then $\frac{dN_2}{dt} = 0$, so

$$n_n N_1 < \sigma v_1 >= n_n < \sigma v >_2 N_2 + \lambda_\beta N_2$$

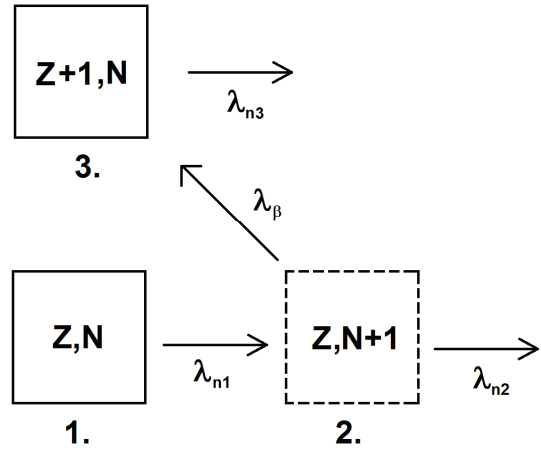

*Fig. 3. The isotonic channel*

If the third nucleus can form only from second nucleus

$$\frac{dN_3}{dt} = \lambda_\beta N_2 - n_n N_3 <\sigma v>_3$$

Beacause of the equilibrium $\frac{dN_3}{dt} = 0$ we get

$$\lambda_\beta N_2 = n_n N_3 <\sigma v>_3$$

The two equations

$$n_n N_1 <\sigma v_1> = n_n N_2 <\sigma v>_2 + \lambda_\beta N_2$$

$$\lambda_\beta N_2 = n_n N_3 <\sigma v>_3$$

about these

$$\frac{N_1 <\sigma v>_1}{N_3 <\sigma v>_3} = \frac{n_n <\sigma v>_2 + \lambda_\beta}{\lambda_\beta}$$

$$\frac{N_1 <\sigma v>_1}{N_3 <\sigma v>_3}\lambda_\beta - \lambda_\beta = n_n <\sigma v>_2$$

From this formula, we can get the neutron density value for equilibrium nucleosynthesis.

$$n_n = \frac{\dfrac{N_1 <\sigma v>_1}{N_3 <\sigma v>_3}\lambda_\beta - \lambda_\beta}{<\sigma v>_2} \text{ or } n_n = \frac{\dfrac{N_1 \; \sigma_1}{N_3 \; \sigma_3} - 1}{\sigma_2}\frac{\lambda_\beta}{v}$$

Because of $<\sigma v> = \sigma \cdot v = $ constant  it is enough to take the neutron capture values at 30keV.

(See the required neutron densities at different stellar temperature in Appendix 3 and 4.)

## 2.3 The role of neutron density

We have the possible equilibrium neutron density at isotopic and isotonic cases as well. There is a big difference between the two cases. In isotopic case the increment of neutron density increases the amount of the third nuclei, but in isotonic case the increment of neutron density decreases the amount of the third nuclei. These result the shift of capture path (or capture band) toward neutron rich nuclei.

## 3. Partial formation of nuclei

We can see in the previous section that not all of the third nuclei can form from the first nuclei. But how much can be formed on this way?

Suppose that from the abundance of the first nuclei ($N_1$) the k-times the abundance of third nuclei ($k \cdot N_3$) are formed ($0 < k$). What neutron density is required in this case? (Here k is the Partial Formation Ratio so PFR.) If $k = 1$ then all of the third nuclei are formed as in the previous section 3.

The required neutron density at the two channels in isotopic and in isotonic cases:

$$n_n = \frac{\lambda_\beta}{v} \frac{\sigma_3}{\sigma_2} \frac{1}{\left(\dfrac{N_1 \sigma_1}{k N_3} - \sigma_3\right)} \qquad\qquad n_n = \frac{\lambda_\beta}{v \sigma_2}\left(\frac{N_1}{N_3} \frac{\sigma_1}{\sigma_3} \frac{1}{k} - 1\right)$$

## 4. Given neutron density

After this theoretic investigation it is necessary to take a realistic approach.

We can consider the inverse question: what part of third nuclei abundance is formed from the first nuclei abundance at given neutron density? ($k = ?$)

### 4.1 In isotopic case:

$$k_Z = R \frac{\lambda_n}{(\lambda_n + \lambda_\beta)} = \frac{N_1 \sigma_1}{N_3 \sigma_3} \frac{\lambda_n}{(\lambda_n + \lambda_\beta)} = \frac{N_1 \sigma_1}{N_3 \sigma_3} \frac{\dfrac{\lambda_n}{\lambda_\beta}}{\left(\dfrac{\lambda_n}{\lambda_\beta} + 1\right)}$$

$$g(x)_Z = R \frac{x}{(x+1)} = \frac{N_1 \sigma_1}{N_3 \sigma_3} \frac{x}{(x+1)} = R \cdot f_n(x) \qquad x = \frac{\lambda_n}{\lambda_\beta}$$

### 4.2 In isotonic case:

$$k_N = R \frac{\lambda_\beta}{(\lambda_n + \lambda_\beta)} = \frac{N_1 \sigma_1}{N_3 \sigma_3} \frac{\lambda_\beta}{(\lambda_n + \lambda_\beta)} = \frac{N_1 \sigma_1}{N_3 \sigma_3} \frac{1}{\left(\dfrac{\lambda_n}{\lambda_\beta} + 1\right)}$$

$$g(x)_N = R \frac{1}{(x+1)} = \frac{N_1 \sigma_1}{N_3 \sigma_3} \frac{1}{(x+1)} = R \cdot f_\beta(x) \qquad x = \frac{\lambda_n}{\lambda_\beta}$$

Where $f_n$ and $f_\beta$ are the classical branching factors for neutron capture and beta decay. So we have uniform functions in both isotopic and isotonic cases.

## 5. Mathematical analysis

It might be better to see these functions in logarithmic representation.

On Fig. 4 the two branching functions are: $f_n(x)$ and $f_\beta(x)$, where $\xi = \lg x$ :

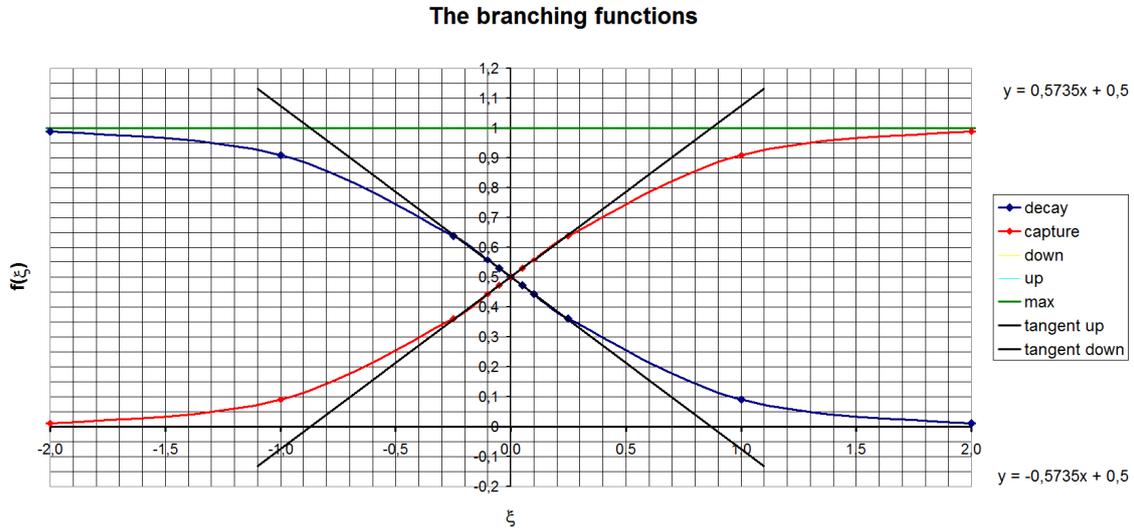

*Fig. 4. The two branching functions in logharitmic scale*

In this logarithmic representation the two functions become symmetric. Then we take the two logarithmic tangents at $\xi = 0$ or $x = 1$.

Where these tangents take the value zero the isotopic formation of nuclei opens, and the isotonic formation of nuclei closes. Similarly, where the tangent takes value one, the isotopic formation is on and the isotonic formation is off. So we have got well-defined characteristics for both of the opening and closing cases.

We found a mathematical definition for the unified interpretation of a branching point closed at the isotonic case and open at the isotopic case.

## 6. The neutron density range

From the mathematical analysis we get the next values:

*Tab. 1.*

| $\xi$ | x | branching ratio | | channel | |
|---|---|---|---|---|---|
| | | $f_n$ | $f_\beta$ | isotopic | isotonic |
| -0.872 | 0,135 | 0.119 | 0.881 | opening | closing |
| 0,872 | 7,398 | 0.881 | 0.119 | opened | closed |

The isotopic channel is beginning to open and isotonic channel is beginning to close when $\lambda_n \approx 0{,}135 \cdot \lambda_\beta$. Isotonic channel is nearly closed and isotopic channel is almost fully opened when $\lambda_n \approx 7{,}39 \cdot \lambda_\beta$. We have reviewed the appropriate three nuclei cases and examined both of $T_s$ and $3T_s$ cases.

At these cases we have got the maximum of all neutron density not more than $n_n = 10^{14}\,\text{cm}^{-3}$. Some details are shown on the Fig. 5. Upper lines refer to isotonic cases; the bottom lines refer to isotonic cases. The temperature dependence of half-lives we took from [4].

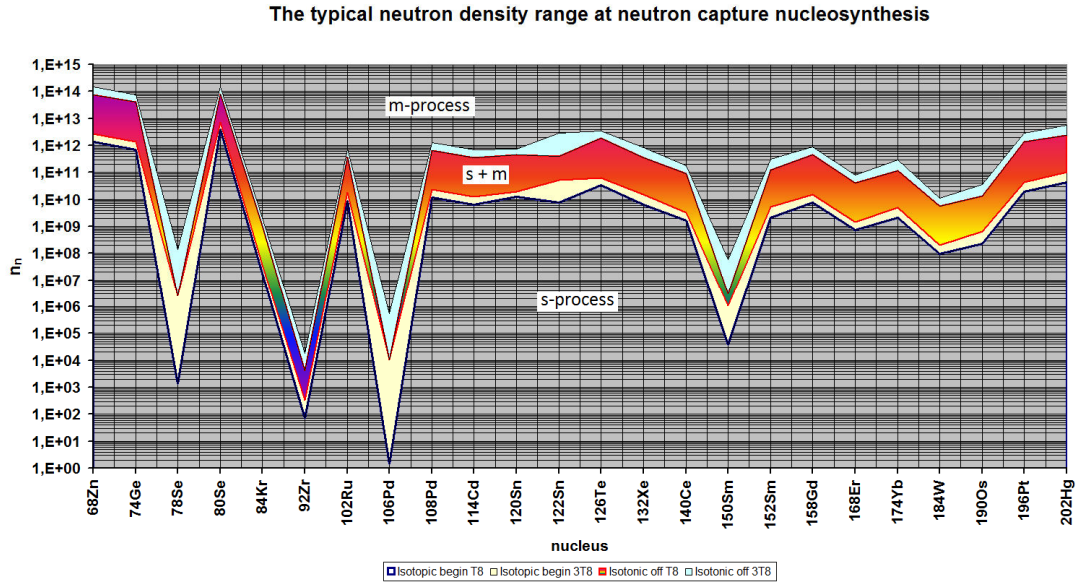

*Fig. 5. The range of neutron capture nucleosynthesis about neutron density.*

All nuclei formation is available at a neutron density range between s-process and r-process. This process is m-process (i-process) (medium or interMediate) [7,8,9,10,11].

## 7. Branching in a new point of view

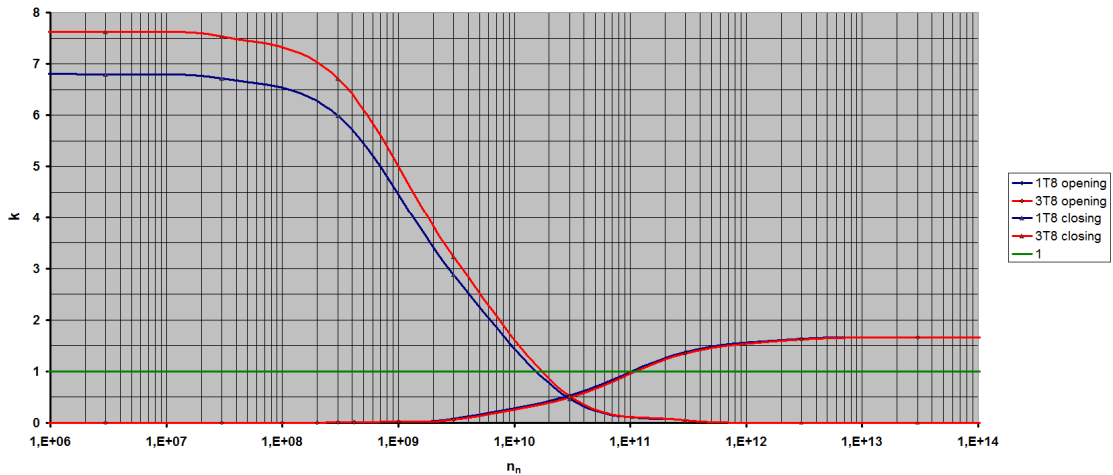

*Fig. 6. Partial formation ratio from $^{102}\text{Ru}$ as the function of neutron density. Here we can the temperature dependence of the formation ratio as well. At $^{102}\text{Ru}$ there is no isomer.*

The branching factor does not give the correct formation ratio. It gives the partial formation ratio: $k = R \cdot f$. (Data from KADoNiS [12], NNDC [13].) The k is determined by f and R, where $R = \dfrac{N_1 \sigma_1}{N_3 \sigma_3}$. The PFR is changing the amount of formed nuclei. The graph on Fig. 6 shows the k as a function of neutron density in the cases of $^{102}Ru \rightarrow ^{103}Rh$ and $^{102}Ru \rightarrow ^{104}Ru$ as well. These are both simple cases with only one incoming channel.

## 8. Full network model

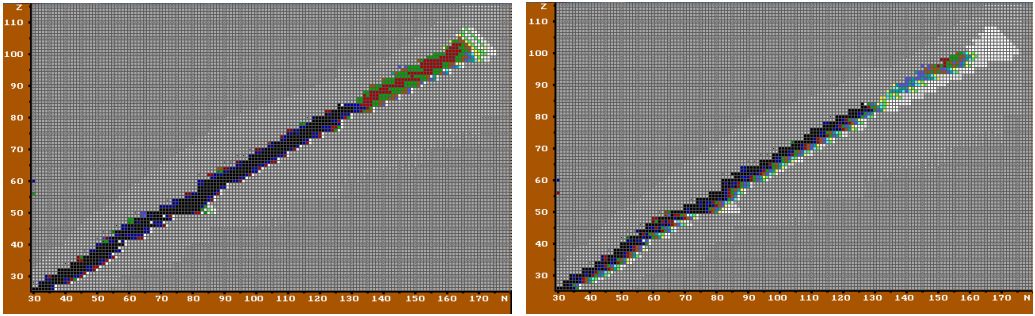

*Fig. 7. The band of neutroncapture nucleosynthesis at AGP TP phase and at IP after TP*

We can see in one unified model the entire possible neutron capture and decay processes [11,14]. Here we can change the neutron density and other parameters. But more data is required here [5]. The formation of nuclei rather occurs along a band than along a path (see Fig. 7). The neutron capture band is only visible in logarithmic representation. The structure of band at $Z = 50$, at tin isotopes is visible on Fig. 8.

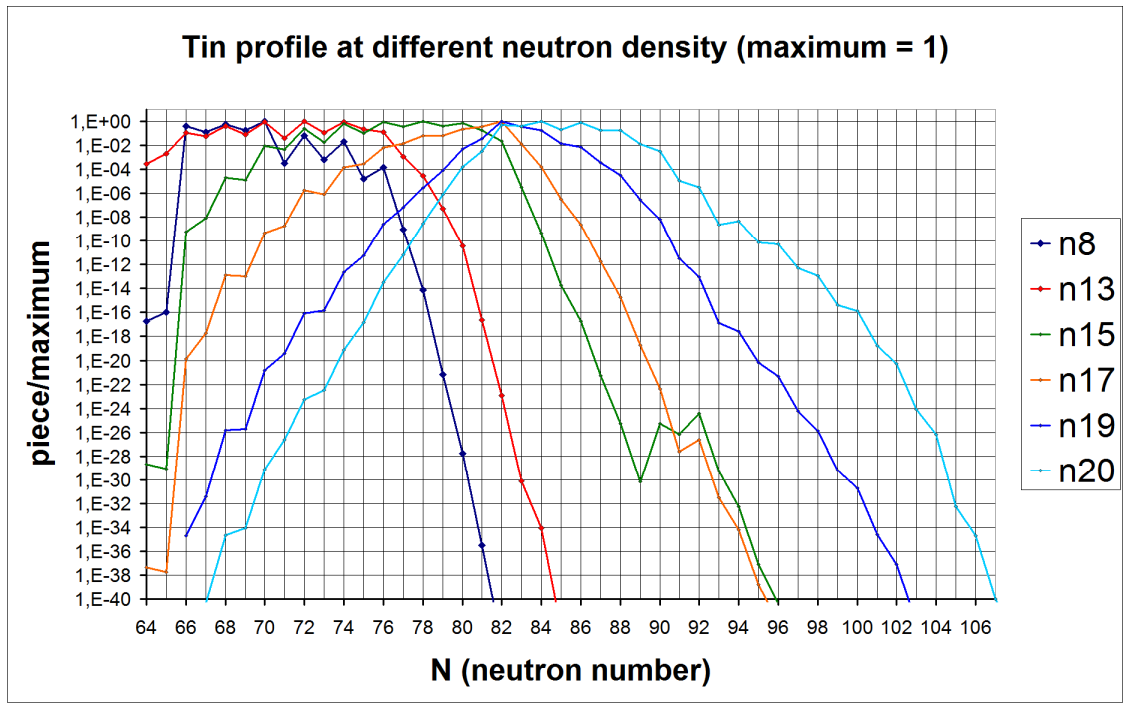

*Fig. 8. The profil of neutron capture band at tin isotopes.* ( $n8 = 10^8 \text{cm}^{-3}$ )

## 9. What does path mean at arbitrary neutron density?

**Paths σ*N max**

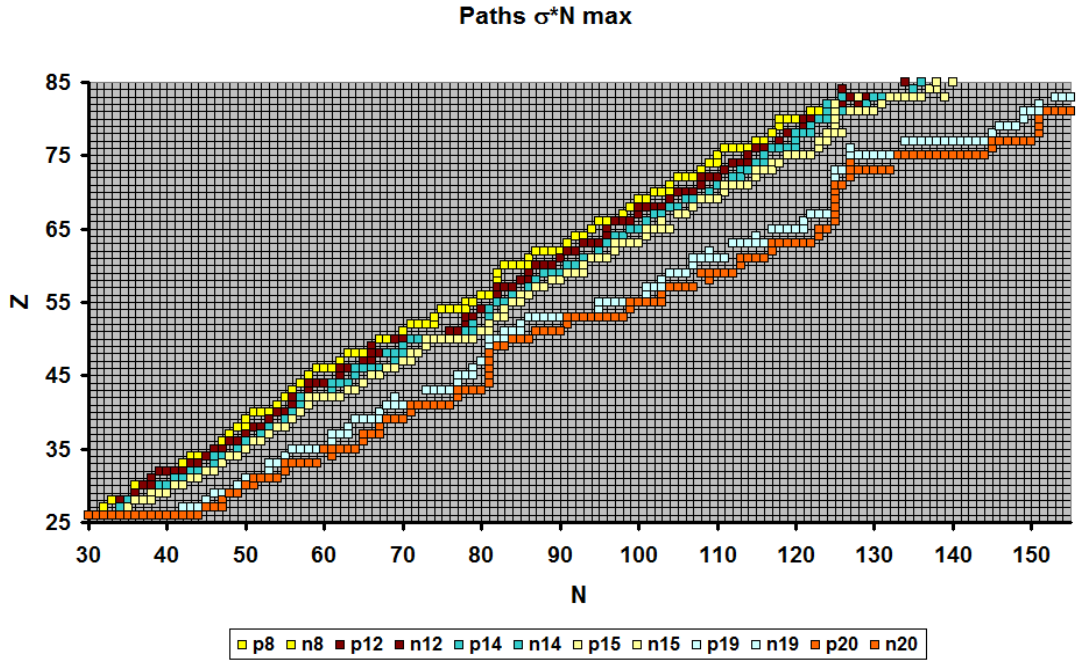

*Fig. 9. The different paths at different neutron density. The maximum values we took from isotonically (p) or isotopically (n). Here p8 and n8 refer to the* $n8 = 10^8 \, \mathrm{cm}^{-3}$ *neutron density.*

Instead of the classical paths the paths are more like the ridge of the $\sigma N$ value at given neutron density, in case of the neutron capture nucleosynthesis. There is s-path, r-path and between them are the m-paths on the Fig. 9 at different neutron density by the unified model [14]. The s-path is the theoretical edge of the paths at very low neutron density.

## 10. Experimental constraints: tellurium and technetium and iron

The isotope anomalies at tellurium [11] and xenon [15] or the presence or absence of the technetium in AGB stars [16,17] can be explained with m-process. These depend on where the „path" is. We have found that the two "r-only" isotopes of tellurium ($^{128}\mathrm{Te}$, $^{130}\mathrm{Te}$) are formed mainly in the m-process at AGB conditions [11].

The existence of $^{60}\mathrm{Fe}$ radioisotope is also important. The stable $^{58}\mathrm{Fe}$ is separated from $^{60}\mathrm{Fe}$ by the unstable isotope $^{59}\mathrm{Fe}$. The case is the same on the Fig. 2. The third nuclei are $^{60}\mathrm{Fe}$. The rate equation for the $^{59}\mathrm{Fe}$:

$$\frac{dN_2}{dt} = n_n N_1 < \sigma v >_1 - n_n N_2 < \sigma v >_2 - \lambda_{\beta 2} N_2$$

$$\frac{dN_3}{dt} = n_n N_2 < \sigma v >_2 - n_n N_3 < \sigma v >_3 - \lambda_{\beta 3} N_3$$

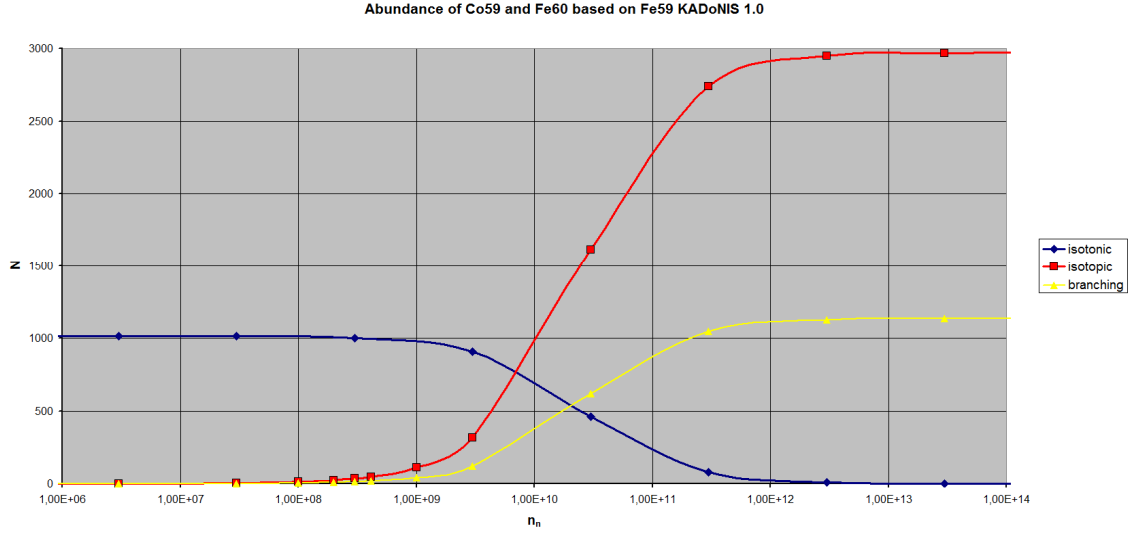

*Fig. 10. The $^{59}$Co and $^{60}$Fe abundance at different neutrondensity. The figure shows the difference beetven PFR and the branchig factor*

In the case of equilibrium comes:

$$N_3 = \frac{N_1 \sigma_1}{\sigma_3 + \dfrac{\lambda_{\beta 3}}{n_n v}} \frac{\lambda_{n2}}{\lambda_{n2} + \lambda_{\beta 2}} = \frac{N_1 \sigma_1}{\sigma_3 + \dfrac{\lambda_{\beta 3}}{n_n v}} f_{n2} \qquad\qquad N_3 = \frac{N_1 \sigma_1}{\sigma_3 + \dfrac{\lambda_{\beta 3}}{n_n v}} \frac{x_2}{1 + x_2}$$

Because of the half-life of $^{60}$Fe ( $T = 1{,}5 \cdot 10^6$ y ) $\dfrac{\lambda_{\beta 3}}{n_n v} \ll \sigma_3$, so it is negligible [18,19].

$$N_3 = \frac{N_1 \sigma_1}{\sigma_3 + \dfrac{\lambda_{\beta 3}}{n_n v}} \frac{x_2}{1 + x_2} \cong \frac{N_1 \sigma_1}{\sigma_3} \frac{x_2}{1 + x_2} = \frac{N_1 \sigma_1}{\sigma_3} f_{n2} = N_1 \frac{\sigma_1}{\sigma_3} f_{n2} = N_1 \frac{13{,}5}{5{,}15} f_{n2} = N_1 \cdot 2{,}62 \cdot f_{n2}$$

At a given neutron density the amount of $^{60}$Fe is nearly three times what we get from the simple branching ratio (see Fig. 10). We used data from KADoNiS 1.0 [19].

This explains the high abundance of $^{60}$Ni, which is four times (4.34 times) the amount of $^{59}$Co. The formation of $^{60}$Ni nuclei occurs mainly through the $^{60}$Fe channel.

## 11. Conclusion

All experienced isotope ratios can be obtained both at $10^8$ K temperature and at $3 \cdot 10^8$ K temperature at intermediate neutron density ($10^{12} \text{-} 10^{14} \text{cm}^{-3}$), so the m-process and the AGB stars are probably one of the main places of nucleosynthesis. It seems that the so-called r-nuclei can form in intermediate processes as well.

## Appendix

## 1. Isotopic equilibrium neutron density at a temperature of $10^8$ K

| | nuc$_1$ | log y$_1$ (Si=6) | σ$_1$ (mb) | nuc$_2$ | T$_2$ | time unit | σ$_2$ (mb) | nuc$_3$ | σ$_3$ (mb) | log y$_3$ (Si=6) | n$_n$ (cm$^{-3}$) | lg n$_n$ |
|---|---|---|---|---|---|---|---|---|---|---|---|---|
| 1. | 62Ni | 3,255 | **22,2** | 63Ni | 100,1 | Y | **66,7** | 64Ni | **8,4** | 2,726 | 2,2E+06 | 6,33 |
| 2. | 68Zn | 2,369 | **20,7** | 69Zn | 56,4 | m | <u>75,4</u> | 70Zn | **10,9** | 0,893 | 1,8E+11 | 11,25 |
| 3. | 68Zn | 2,369 | **20,7** | 69Zn$^m$ | 13,76 | d* | *110,9* | 70Zn | **10,9** | 0,893 | 3,5E+08 | 8,54 |
| 4. | 69Ga | 1,358 | **118,7** | 70Ga | 21,14 | d | <u>301,5</u> | 71Ga | ***104,6*** | 1,176 | 6,4E+09 | 9,81 |
| 5. | 74Ge | 1,638 | **37,6** | 75Ge | 82,78 | m | *203,2* | *76Ge* | 24,4 | 0,965 | 4,0E+11 | 11,61 |
| 6. | 78Se | 1,164 | **61,1** | 79Se | 2,95E+05 | y | ***263*** | 80Se | 38 | 1,49 | -4,3E+04 | n/a |
| 7. | 80Se | 1,49 | **38** | 81Se | 18,45 | m | <u>82,3</u> | *82Se* | 8,4 | 0,757 | 1,2E+12 | 12,08 |
| 8. | 80Se | 1,49 | **38** | 81Se$^m$ | 57,28 | m* | <u>82,3</u> | *82Se* | 8,4 | 0,757 | 3,8E+11 | 11,59 |
| 9. | 79Br | 0,775 | **622** | 80Br | 1,77E+01 | m | <u>790,0</u> | 81Br | 239 | 0,766 | 1,8E+12 | 12,25 |
| 10. | 79Br | 0,775 | **622** | 80Br$^m$ | 4,42E+00 | h* | <u>790,0</u> | 81Br | 239 | 0,766 | 1,2E+11 | 11,08 |
| 11. | 84Kr | 1,408 | **32,6** | 85Kr | 3916,8 | D | **73** | 86Kr | **4,76** | 0,893 | 4,8E+06 | 6,68 |
| 12. | 84Kr | 1,408 | **32,6** | 85Kr$^m$ | 4,48 | h 78,6% | **73** | 86Kr | **4,76** | 0,893 | 1,0E+11 | 11,01 |
| 13. | 85Rb | 0,709 | **234** | 86Rb | 18,642 | d | ***202*** | 87Rb | **15,7** | 0,316 | 2,2E+08 | 8,34 |
| 14. | 92Zr | 0,209 | **37,8** | 93Zr | 1,53E+06 | Y | **96** | *94Zr* | 27,3 | 0,209 | 1,3E+03 | 3,12 |
| 15. | 94Zr | 0,297 | **27,3** | 95Zr | 64,032 | D | **79** | *96Zr* | 10,3 | -0,496 | 3,8E+08 | 8,58 |
| 16. | 98Mo | -0,218 | **70,5** | 99Mo | 2,7489 | D | ***240*** | *100Mo* | 80,5 | -0,609 | 3,9E+10 | 10,59 |
| 17. | 102Ru | -0,231 | **151** | 103Ru | 29,26 | D | ***343*** | *104Ru* | 154 | -0,461 | 9,5E+10 | 10,98 |
| 18. | 106Pd | -0,42 | **244** | 107Pd | 6,50E+06 | Y | **1302** | 108Pd | 218 | -0,431 | 7,4E+01 | 1,87 |
| 19. | 108Pd | -0,431 | **218** | 109Pd | 13,70012 | H | *236,3* | *110Pd* | 157 | -0,785 | 1,0E+11 | 11,01 |
| 20. | 107Ag | -0,602 | **787** | 108Ag | 2,37 | m | <u>1788,0</u> | 109Ag | 793 | -0,627 | 2,0E+14 | 14,29 |
| 21. | 107Ag | -0,602 | **787** | 108Ag$^m$ | 438 | y*ec | *1383,0* | 109Ag | 793 | -0,627 | 2,6E+06 | 6,42 |
| 22. | 114Cd | -0,333 | **135,3** | 115Cd | 53,46 | H | ***290*** | *116Cd* | 76,1 | -0,914 | 7,9E+09 | 9,90 |
| 23. | 114Cd | -0,333 | **135,3** | 115Cd$^m$ | 44,56 | d | 224 | *116Cd* | 76,1 | -0,914 | 5,1E+08 | 8,71 |
| 24. | 113In | -2,103 | **809** | 114In | 71,9 | s | <u>1308,0</u> | 115In | **776** | -0,754 | -2,8E+13 | n/a |
| 25. | 113In | -2,103 | **809** | 114In$^m$ | 49,51 | d* | 2595 | 115In | **776** | -0,754 | -2,4E+08 | n/a |
| 26. | 120Sn | 0,097 | **36,3** | 121Sn | 27,03 | H | 167 | *122Sn* | 22,6 | -0,745 | 9,1E+09 | 9,96 |
| 27. | 120Sn | 0,097 | **36,3** | 121Sn$^m$ | 43,9 | y 22% | *175,4* | *122Sn* | 22,6 | -0,745 | 6,1E+05 | 5,79 |
| 28. | 122Sn | -0,745 | **22,6** | 123Sn | 129,2 | D | 361 | *124Sn* | 15,7 | -0,644 | 3,9E+11 | 11,59 |
| 29. | 122Sn | -0,745 | **22,6** | 123Sn$^m$ | 40,06 | m | 361 | *124Sn* | 15,7 | -0,644 | 1,8E+15 | 15,26 |
| 30. | 121Sb | -0,752 | **532** | 122Sb | **3** | **D** | 894 | 123Sb | 303 | -0,879 | 9,0E+09 | 9,95 |
| 31. | 126Te | -0,046 | **81,3** | 127Te | 9,35 | H | *256,8* | *128Te* | 44,4 | 0,185 | 3,9E+12 | 12,59 |
| 32. | 126Te | -0,046 | **81,3** | 127Te$^m$ | 109 | d 2,4% | *668,6* | *128Te* | 44,4 | 0,185 | 1,1E+11 | 11,05 |
| 33. | 128Te | 0,185 | **44,4** | 129Te | 69,6 | M | <u>137,5</u> | *130Te* | **14,2** | 0,22 | 2,4E+12 | 12,37 |

| | nuc₁ | log y₁ (Si=6) | σ₁ (mb) | nuc₂ | T₂ | time unit | σ₂ (mb) | nuc₃ | σ₃ (mb) | log y₃ (Si=6) | nₙ (cm⁻³) | lg nₙ |
|---|---|---|---|---|---|---|---|---|---|---|---|---|
| 34. | 128Te | 0,185 | **44,4** | 129Teᵐ | 33,6 | d* | *493,6* | *130Te* | **14,2** | 0,22 | 9,5E+08 | 8,98 |
| 35. | 132Xe | 0,086 | **63,8** | 133Xe | 5,243 | D | 127 | *134Xe* | **21,3** | -0,321 | 7,2E+09 | 9,86 |
| 36. | 134Xe | -0,321 | **21,3** | 135Xe | 9,14 | H | 65,6 | *136Xe* | **0,98** | -0,405 | 4,7E+10 | 10,67 |
| 37. | 140Ce | 0,004 | **11,73** | 141Ce | 32,508 | D | 76 | *142Ce* | **29,9** | -0,9 | 5,6E+09 | 9,75 |
| 38. | 146Nd | -0,845 | **91,2** | 147Nd | 10,98 | D | 544 | *148Nd* | **146,6** | -1,324 | 5,7E+09 | 9,75 |
| 39. | 148Nd | -1,324 | **146,6** | 149Nd | 17,28 | H | *513,2* | *150Nd* | **156,1** | -1,333 | -1,9E+12 | n/a |
| 40. | 150Sm | -1,717 | **422,3** | 151Sm | 90 | Y | **3040** | *152Sm* | **464,8** | -1,161 | -3,7E+05 | n/a |
| 41. | 152Sm | -1,161 | **464,8** | 153Sm | 46,284 | H | *1095* | *154Sm* | **216,9** | -1,232 | 1,0E+10 | 10,01 |
| 42. | 158Gd | -1,086 | **323,6** | 159Gd | 18,479 | H | *455,2* | *160Gd* | **178** | -1,141 | 7,6E+10 | 10,88 |
| 43. | 168Er | -1,168 | **319** | 169Er | 9,392 | D | 653,0 | *170Er* | **170,2** | -1,428 | 2,2E+09 | 9,35 |
| 44. | 174Yb | -1,103 | **150,5** | 175Yb | 4,185 | D | 558 | *176Yb* | **115,9** | -1,5 | 6,7E+09 | 9,83 |
| 45. | 184W | -1,389 | **225** | 185W | 75,1 | D | **633** | *186W* | **226** | -1,423 | 8,8E+09 | 9,94 |
| 46. | 185Re | -1,717 | **1438,5** | 186Re | 3,7186 | D | **743** | 187Re | **1184** | -1,475 | -3,5E+10 | n/a |
| 47. | 185Re | -1,717 | **1438,5** | 186Reᵐ | 2,00E+05 | y* | **743** | 187Re | **1184** | -1,475 | -1,8E+03 | n/a |
| 48. | 190Os | -0,75 | **278** | 191Os | 15,4 | D | *1290* | *192Os* | **160** | -0,558 | 1,4E+10 | 10,16 |
| 49. | 191Ir | -0,607 | **1350** | 192Ir | 73,827 | D | *2080* | 193Ir | **994** | -0,383 | -1,0E+09 | n/a |
| 50. | 191Ir | -0,607 | **1350** | 192Irᵐ | 241 | y*ec | *2080* | 193Ir | **994** | -0,383 | -8,6E+05 | n/a |
| 51. | 196Pt | -0,47 | **167,4** | 197Pt | 19,8915 | H | 79,7 | *198Pt* | **94** | -1,015 | 8,6E+10 | 10,93 |
| 52. | 202Hg | -0,996 | **63,3** | 203Hg | 46,594 | D | *98* | *204Hg* | **42** | -1,633 | 5,7E+10 | 10,76 |
| 53. | 203Tl | -1,265 | **170,5** | 204Tl | 3,78 | y | 215 | *205Tl* | **53** | -0,866 | 3,4E+08 | 8,53 |

## 2. Isotopic equilibrium neutron density at a temperature of $3 \cdot 10^8 \, \mathrm{K}$

| | nuc₁ | log y₁ (Si=6) | σ₁ (mb) | nuc₂ | T₂ | time unit | σ₂ (mb) | nuc₃ | σ₃ (mb) | log y₃ (Si=6) | nₙ (cm⁻³) | lg nₙ |
|---|---|---|---|---|---|---|---|---|---|---|---|---|
| 1. | 62Ni | 3,255 | **22,2** | 63Ni | 100,1 | Y | **66,7** | 64Ni | **8,4** | 2,726 | 1,2E+07 | 7,09 |
| 2. | 68Zn | 2,369 | **20,7** | 69Zn | 56,4 | m | <u>75,4</u> | 70Zn | **10,9** | 0,893 | 1,8E+11 | 11,25 |
| 3. | 68Zn | 2,369 | **20,7** | 69Znᵐ | 13,76 | d* | *110,9* | 70Zn | **10,9** | 0,893 | 3,5E+08 | 8,54 |
| 4. | 69Ga | 1,358 | **118,7** | 70Ga | 21,14 | d | <u>301,5</u> | 71Ga | **104,6** | 1,176 | 6,4E+09 | 9,81 |
| 5. | 74Ge | 1,638 | **37,6** | 75Ge | 82,78 | m | *203,2* | *76Ge* | **24,4** | 0,965 | 4,0E+11 | 11,60 |
| 6. | 78Se | 1,164 | **61,1** | 79Se | 2,95E+05 | y | 263 | 80Se | 38 | 1,49 | -7,6E+07 | n/a |
| 7. | 80Se | 1,49 | **38** | 81Se | 18,45 | m | <u>82,3</u> | *82Se* | **8,4** | 0,757 | 1,1E+12 | 12,05 |
| 8. | 80Se | 1,49 | **38** | 81Seᵐ | 57,28 | m* | 82,3 | *82Se* | **8,4** | 0,757 | 3,6E+11 | 11,55 |
| 9. | 79Br | 0,775 | **622** | 80Br | 1,77E+01 | m | <u>790,0</u> | 81Br | 239 | 0,766 | 1,2E+12 | 12,08 |
| 10. | 79Br | 0,775 | **622** | 80Brᵐ | 4,42E+00 | h* | <u>790,0</u> | 81Br | 239 | 0,766 | 7,9E+10 | 10,90 |
| 11. | 84Kr | 1,408 | **32,6** | 85Kr | 3916,8 | D | **73** | 86Kr | **4,76** | 0,893 | 4,9E+06 | 6,69 |

| | nuc$_1$ | log y$_1$ (Si=6) | σ$_1$ (mb) | nuc$_2$ | T$_2$ | time unit | σ$_2$ (mb) | nuc$_3$ | σ$_3$ (mb) | log y$_3$ (Si=6) | n$_n$ (cm$^{-3}$) | lg n$_n$ |
|---|---|---|---|---|---|---|---|---|---|---|---|---|
| 12. | 84Kr | 1,408 | **32,6** | 85Kr$^m$ | 4,48 | h 78,6% | **73** | 86Kr | **4,76** | 0,893 | 1,0E+11 | 11,02 |
| 13. | 85Rb | 0,709 | **234** | 86Rb | 18,642 | d | 202 | 87Rb | **15,7** | 0,316 | 2,3E+11 | 11,36 |
| 14. | 92Zr | 0,209 | **37,8** | 93Zr | 1,53E+06 | Y | **96** | *94Zr* | 27,3 | 0,209 | 4,8E+03 | 3,68 |
| 15. | 94Zr | 0,297 | **27,3** | 95Zr | 64,032 | D | 79 | *96Zr* | 10,3 | -0,496 | 3,8E+08 | 8,58 |
| 16. | 98Mo | -0,218 | **70,5** | 99Mo | 2,7489 | D | 240 | *100Mo* | 80,5 | -0,609 | 5,8E+10 | 10,76 |
| 17. | 102Ru | -0,231 | **151** | 103Ru | 29,26 | D | 343 | *104Ru* | **154** | -0,461 | 1,1E+11 | 11,03 |
| 18. | 106Pd | -0,42 | **244** | 107Pd | 6,50E+06 | Y | **1302** | 108Pd | **218** | -0,431 | 5,2E+05 | 5,72 |
| 19. | 108Pd | -0,431 | **218** | 109Pd | 13,70012 | H | *236,3* | *110Pd* | 157 | -0,785 | 1,0E+11 | 11,01 |
| 20. | 107Ag | -0,602 | **787** | 108Ag | 2,37 | m | <u>1788,0</u> | 109Ag | 793 | -0,627 | 1,7E+14 | 14,23 |
| 21. | 107Ag | -0,602 | **787** | 108Ag$^m$ | 438 | y*ec | *1383,0* | 109Ag | 793 | -0,627 | 1,0E+06 | 6,36 |
| 22. | 114Cd | -0,333 | **135,3** | 115Cd | 53,46 | H | 290 | *116Cd* | 76,1 | -0,914 | 7,9E+09 | 9,90 |
| 23. | 114Cd | -0,333 | **135,3** | 115Cd$^m$ | 44,56 | d | 224 | *116Cd* | 76,1 | -0,914 | 5,1E+08 | 8,71 |
| 24. | 113In | -2,103 | **809** | 114In | 71,9 | s | <u>1308,0</u> | 115In | 776 | -0,754 | -2,8E+13 | n/a |
| 25. | 113In | -2,103 | **809** | 114In$^m$ | 49,51 | d* | 2595 | 115In | 776 | -0,754 | -2,4E+08 | n/a |
| 26. | 120Sn | 0,097 | **36,3** | 121Sn | 27,03 | H | 167 | *122Sn* | 22,6 | -0,745 | 4,6E+09 | 9,66 |
| 27. | 120Sn | 0,097 | **36,3** | 121Sn$^m$ | 43,9 | y 22% | *175,4* | *122Sn* | 22,6 | -0,745 | 3,0E+05 | 5,48 |
| 28. | 122Sn | -0,745 | **22,6** | 123Sn | 129,2 | D | 361 | *124Sn* | 15,7 | -0,644 | 2,4E+12 | 12,38 |
| 29. | 122Sn | -0,745 | **22,6** | 123Sn$^m$ | 40,06 | m | 361 | *124Sn* | 15,7 | -0,644 | 1,1E+16 | 16,04 |
| 30. | 121Sb | -0,752 | **532** | 122Sb | **3** | D | 894 | 123Sb | **303** | -0,879 | 3,4E+10 | 10,53 |
| 31. | 126Te | -0,046 | **81,3** | 127Te | 9,35 | H | *256,8* | *128Te* | 44,4 | 0,185 | 3,4E+12 | 12,53 |
| 32. | 126Te | -0,046 | **81,3** | 127Te$^m$ | 109 | d 2,4% | *668,6* | *128Te* | 44,4 | 0,185 | 9,6E+10 | 10,98 |
| 33. | 128Te | 0,185 | **44,4** | 129Te | 69,6 | M | <u>137,5</u> | *130Te* | 14,2 | 0,22 | 2,2E+12 | 12,35 |
| 34. | 128Te | 0,185 | **44,4** | 129Te$^m$ | 33,6 | d* | *493,6* | *130Te* | 14,2 | 0,22 | 9,0E+08 | 8,95 |
| 35. | 132Xe | 0,086 | **63,8** | 133Xe | 5,243 | D | 127 | *134Xe* | 21,3 | -0,321 | 9,4E+09 | 9,97 |
| 36. | 134Xe | -0,321 | **21,3** | 135Xe | 9,14 | H | 65,6 | *136Xe* | **0,98** | -0,405 | 4,7E+10 | 10,67 |
| 37. | 140Ce | 0,004 | **11,73** | 141Ce | 32,508 | D | 76 | *142Ce* | 29,9 | -0,9 | 5,6E+09 | 9,75 |
| 38. | 146Nd | -0,845 | **91,2** | 147Nd | 10,98 | D | 544 | *148Nd* | 146,6 | -1,324 | 5,7E+09 | 9,75 |
| 39. | 148Nd | -1,324 | **146,6** | 149Nd | 17,28 | H | *513,2* | *150Nd* | 156,1 | -1,333 | -1,9E+12 | n/a |
| 40. | 150Sm | -1,717 | **422,3** | 151Sm | 90 | Y | **3040** | *152Sm* | 464,8 | -1,161 | -9,8E+06 | n/a |
| 41. | 152Sm | -1,161 | **464,8** | 153Sm | 46,284 | H | 1095 | *154Sm* | 216,9 | -1,232 | 1,6E+10 | 10,20 |
| 42. | 158Gd | -1,086 | **323,6** | 159Gd | 18,479 | H | *455,2* | *160Gd* | 178 | -1,141 | 7,6E+10 | 10,88 |
| 43. | 168Er | -1,168 | **319** | 169Er | 9,392 | D | 653,0 | *170Er* | 170,2 | -1,428 | 2,2E+09 | 9,33 |
| 44. | 174Yb | -1,103 | **150,5** | 175Yb | 4,185 | D | 558 | *176Yb* | 115,9 | -1,5 | 1,0E+10 | 10,00 |
| 45. | 184W | -1,389 | **225** | 185W | 75,1 | D | **633** | *186W* | 226 | -1,423 | 9,5E+09 | 9,98 |
| 46. | 185Re | -1,717 | **1438,5** | 186Re | 3,7186 | D | **743** | 187Re | 1184 | -1,475 | -3,8E+10 | n/a |
| 47. | 185Re | -1,717 | **1438,5** | 186Re$^m$ | 2,00E+05 | y*ec | **743** | 187Re | 1184 | -1,475 | -1,9E+03 | n/a |
| 48. | 190Os | -0,75 | **278** | 191Os | 15,4 | D | 1290 | *192Os* | 160 | -0,558 | 2,6E+10 | 10,41 |

|      | nuc$_1$ | log y$_1$ (Si=6) | σ$_1$ (mb) | nuc$_2$ | T$_2$ | time unit | σ$_2$(mb) | nuc$_3$ | σ$_3$ (mb) | log y$_3$ (Si=6) | n$_n$ (cm$^{-3}$) | lg n$_n$ |
|------|---------|--------|---------|---------|--------|------|---------|---------|---------|--------|---------|---------|
| 49. | 191Ir | -0,607 | **1350** | 192Ir | 73,827 | D | 2080 | 193Ir | **994** | -0,383 | -1,4E+09 | n/a |
| 50. | 191Ir | -0,607 | **1350** | 192Ir$^m$ | 2,41E+02 | y* | 2080 | 193Ir | **994** | -0,383 | -1,2E+06 | n/a |
| 51. | 196Pt | -0,47 | **167,4** | 197Pt | 19,8915 | H | *79,7* | *198Pt* | **94** | -1,015 | 9,9E+10 | 11,00 |
| 52. | 202Hg | -0,996 | **63,3** | 203Hg | 46,594 | D | 98 | *204Hg* | **42** | -1,633 | 8,1E+10 | 10,91 |
| 53. | 203Tl | -1,265 | **170,5** | 204Tl | 3,78 | y | 215 | 205Tl | **53** | -0,866 | 6,8E+10 | 10,83 |

## 3. Isotonic equilibrium neutron density at a temperature of $10^8$ K

|      | nuc$_1$ | log y$_1$ (Si=6) | σ$_1$ (mb) | nuc$_2$ | T$_2$ | time unit | σ$_2$(mb) | nuc$_3$ | σ$_3$ (mb) | log y$_3$ (Si=6) | n$_n$ (cm$^{-3}$) | lg n$_n$ |
|------|---------|--------|---------|---------|--------|------|---------|---------|---------|--------|---------|---------|
| 1. | 58Fe | 3,473 | **13,3** | 44,4495 | d | | 59Fe | **26,4** | 59Co | **40,1** | 3,352 | -1,4E+10 | n/a |
| 2. | 59Co | 3,352 | **40,1** | 1925,28 | d | | 60Co | *12,64* | 60Ni | **29,9** | 4,11 | -1,2E+09 | n/a |
| 3. | 62Ni | 3,255 | **22,2** | 100,1 | y | | 63Ni | **66,7** | 63Cu | **53,7** | 2,558 | 1,8E+07 | 7,26 |
| 4. | *64Ni* | *2,726* | **8** | *2,5172* | *h* | | 65Ni | 21,2 | *65Cu* | **29** | 2,207 | -1,2E+12 | n/a |
| 5. | *65Cu* | *2,207* | **29** | *5,12* | *m* | | 66Cu | 126,6 | *66Zn* | **36,4** | 2,544 | -4,2E+13 | n/a |
| 6. | 68Zn | 2,369 | **20,7** | 56,4 | m | | 69Zn | 75,4 | 69Ga | **118,7** | 1,358 | 7,9E+12 | 12,90 |
| 7. | 68Zn | 2,369 | **20,7** | 13,76 | h* | | 69Zn$^m$ | *110,9* | 69Ga | **118,7** | 1,358 | 3,7E+11 | 11,56 |
| 8. | *70Zn* | *0,893* | **10,9** | *2,45* | *m* | | 71Zn | 48,26 | *71Ga* | **106,4** | 1,176 | -3,4E+14 | n/a |
| 9. | *70Zn* | *0,893* | **10,9** | *3,96* | *h* | | 71Zn$^m$ | 48,26 | *71Ga* | **106,4** | 1,176 | -3,5E+12 | n/a |
| 10. | 69Ga | 1,358 | **118,7** | 21,14 | m | | 70Ga | 301,5 | 70Ge | **89,1** | 1,387 | 1,6E+12 | 12,22 |
| 11. | 71Ga | 1,176 | **106,4** | 14,095 | h | | 72Ga | *267,6* | 72Ge | **53,1** | 1,513 | -1,7E+10 | n/a |
| 12. | 74Ge | 1,638 | **37,4** | 82,78 | m | | 75Ge | *203,1* | 75As | **355** | 0,817 | -7,7E+11 | n/a |
| 13. | 75As | 0,817 | **355** | 1,0942 | d | | 76As | *469,6* | 76Se | **168** | 0,748 | 4,9E+11 | 11,69 |
| 14. | 78Se | 1,164 | **61,1** | 2,95E+05 | y | | 79Se | 263 | 79Br | **622** | 0,775 | -7,8E+03 | n/a |
| 15. | 80Se | 1,49 | **38** | 18,45 | m | | 81Se | 229 | 81Br | **239** | 0,766 | -1,6E+12 | n/a |
| 16. | 80Se | 1,49 | **38** | 57,28 | m* | | 81Se$^m$ | 229 | 81Br | **239** | 0,766 | -5,1E+11 | n/a |
| 17. | 79Br | 0,775 | **622** | 17,68 | m | | 80Br | 790 | 80Kr | **274** | 0,009 | 3,6E+13 | 13,56 |
| 18. | 79Br | 0,775 | **622** | 4,4205 | h* | | 80Br$^m$ | 790 | 80Kr | **90,4** | 0,009 | 7,8E+12 | 12,89 |
| 19. | 81Br | 0,766 | **239** | 35,282 | h | | 82Br | *390,5* | 82Kr | **93** | 0,716 | 1,0E+11 | 11,00 |
| 20. | 84Kr | 1,408 | **32,6** | 3916,8 | d | | 85Kr | 73 | 85Rb | **234** | 0,709 | -3,1E+07 | n/a |
| 21. | 84Kr | 1,408 | **32,6** | 4,78 | h79% | | 85Kr$^m$ | 73 | 85Rb | **234** | 0,709 | -6,2E+11 | n/a |
| 22. | 86Kr | 0,893 | **4,76** | 76,3 | m | | 87Kr | 18 | 87Rb | **15,7** | 0,316 | 4,5E+12 | 12,65 |
| 23. | 85Rb | 0,709 | **234** | 18,642 | d | | 86Rb | 202 | 86Sr | **63,5** | 0,365 | 5,6E+10 | 10,75 |
| 24. | 87Rb | 0,316 | **15,7** | 17,773 | m | | 88Rb | 110 | 88Sr | **6,16** | 1,288 | -1,5E+13 | n/a |
| 25. | 88Sr | 1,288 | **6,16** | 50,57 | d | | 89Sr | 19 | 89Y | **19,3** | 0,667 | 1,0E+10 | 10,01 |

| | nuc$_1$ | log y$_1$ (Si=6) | σ$_1$(mb) | nuc$_2$ | T$_2$ | time unit | σ$_2$(mb) | nuc$_3$ | σ$_3$(mb) | log y$_3$ (Si=6) | n$_n$ (cm$^{-3}$) | lg n$_n$ |
|---|---|---|---|---|---|---|---|---|---|---|---|---|
| 26. | 89Y | 0,667 | **19,3** | 90 | y | 90Y | *149,9* | 90Zr | **19,3** | 0,769 | -1,3E+06 | n/a |
| 27. | 92Zr | 0,299 | **37,8** | 1,53E+06 | y | 93Zr | 96 | 93Nb | **265,7** | 0,299 | -4,4E+02 | n/a |
| 28. | 93Nb | -0,156 | **265,7** | 2,03E+04 | y | 94Nb | 482 | 94Mo | **109,6** | -0,636 | 2,0E+09 | 9,30 |
| 29. | 102Ru | -0,231 | **151** | 39,26 | d | 103Ru | 343 | 103Rh | **810** | -0,463 | -3,2E+10 | n/a |
| 30. | 103Rh | -0,463 | **810** | 42,3 | s | 104Rh | 154 | 104Pd | **274** | -0,818 | 2,2E+15 | 15,35 |
| 31. | 103Rh | -0,463 | **810** | 4,34 | m* | 104Rh$^m$ | 154 | 104Pd | **274** | -0,818 | 3,6E+14 | 14,56 |
| 32. | 106Pd | -0,42 | **244** | 6,60E+06 | y | 107Pd | 1302 | 107Ag | **787** | -0,602 | -5,7E+00 | n/a |
| 33. | 108Pd | -0,431 | **218** | 13,7012 | h | 109Pd | *236,3* | 109Ag | **793** | -0,627 | -1,2E+11 | n/a |
| 34. | 107Ag | -0,602 | **787** | 2,37 | m | 108Ag | 1788 | 108Cd | **202** | -1,851 | 6,8E+14 | 14,84 |
| 35. | 107Ag | -0,602 | **787** | 438 | y* | 108Ag$^m$ | *1383,0* | 108Cd | **202** | -1,851 | 9,1E+06 | 6,96 |
| 36. | 109Ag | -0,627 | **793** | 24,6 | s | 110Ag | 1172 | 110Cd | **229,9** | -0,701 | 1,1E+14 | 14,05 |
| 37. | 109Ag | -0,627 | **793** | 249,76 | d | 110Ag$^m$ | 1172 | 110Cd | **229,9** | -0,701 | 1,3E+08 | 8,11 |
| 38. | 114Cd | -0,333 | **135,3** | 53,46 | h | 115Cd | 290 | 115In | **776** | -0,754 | -2,5E+10 | n/a |
| 39. | 114Cd | -0,333 | **135,3** | 44,56 | d* | 115Cd$^m$ | 224 | 115In | **776** | -0,754 | -1,6E+09 | n/a |
| 40. | 113In | -2,103 | **229** | 71,9 | s | 114In | 1308 | 114Sn | **134,4** | -1,599 | -1,3E+13 | n/a |
| 41. | 113In | -2,103 | **580** | 49,51 | d* | 114In$^m$ | 2595 | 114Sn | **134,4** | -1,599 | 8,1E+07 | 7,91 |
| 42. | 115In | -0,754 | **154** | 14,1 | s | 116In | 1377 | 116Sn | **92,3** | -0,263 | -6,1E+13 | n/a |
| 43. | 115In | -0,754 | **622** | 54,29 | m | 116In$^m$ | 1377 | 116Sn | **92,3** | -0,263 | 6,7E+11 | 11,83 |
| 44. | 120Sn | 0,097 | **36,3** | 27,03 | h | 121Sn | 167 | 121Sb | **532** | -0,752 | -3,3E+10 | n/a |
| 45. | 120Sn | 0,097 | **36,3** | 43,9 | y* | 121Sn$^m$ | *175,4* | 121Sb | **532** | -0,752 | -2,2E+06 | n/a |
| 46. | 122Sn | -0,745 | **22,6** | 129,2 | d | 123Sn | 361 | 123Sb | **303** | -0,879 | -4,0E+10 | n/a |
| 47. | 122Sn | -0,745 | **21,7** | 40,06 | m | 123Sn$^m$ | 361 | 123Sb | **303** | -0,879 | -1,9E+14 | n/a |
| 48. | 121Sb | -0,752 | **532** | 2,7238 | d | 122Sb | 894 | 122Te | **295,4** | -0,928 | 2,1E+10 | 10,31 |
| 49. | 123Sb | -0,879 | **303** | 60,11 | d | 124Sb | *924,0* | 124Te | **155** | -0,654 | 3,1E+11 | 11,49 |
| 50. | 126Te | -0,046 | **81,3** | 9,35 | h | 127Te | *256,8* | 127I | **662** | -0,046 | -2,6E+11 | n/a |
| 51. | 126Te | -0,046 | **81,3** | 109 | d* | 127Te$^m$ | *668,6* | 127I | **662** | -0,046 | -3,6E+08 | n/a |
| 52. | 127I | -0,046 | **662** | 24,99 | m | 128I | *679,5* | 128Xe | **262,5** | 0,991 | -1,8E+12 | n/a |
| 53. | 132Xe | 0,086 | **63,8** | 5,243 | d | 133Xe | 127 | 133Cs | **502** | -0,429 | -2,7E+10 | n/a |
| 54. | 133Cs | -0,429 | **502** | 2,0652 | y | 134Cs | 724 | 134Ba | **176** | -0,963 | 1,5E+09 | 9,17 |
| 55. | 138Ba | 0,508 | **4,13** | 83,06 | m | 139Ba | *39,7* | 139La | **32,4** | -0,351 | -1,0E+12 | n/a |
| 56. | 139La | -0,351 | **32,4** | 1,67855 | d | 140La | 117,8 | 140Ce | **11,73** | 0,004 | 3,3E+10 | 10,52 |
| 57. | 140Ce | 0,004 | **11,73** | 32,508 | d | 141Ce | 76 | 141Pr | **111,4** | -0,777 | -4,4E+09 | n/a |
| 58. | 141Pr | -0,777 | **111,4** | 19,12 | h | 142Pr | 361,4 | 142Nd | **35,1** | -0,65 | 5,4E+10 | 10,73 |
| 59. | 150Sm | -1,717 | **422,3** | 90 | y | 151Sm | 3040 | 151Eu | **3556** | 0,35 | -2,8E+05 | n/a |
| 60. | 152Sm | -1,161 | **464,8** | 46,284 | h | 153Sm | 1095 | 153Eu | **2567** | -1,294 | -1,2E+10 | n/a |
| 61. | 153Eu | -1,294 | **2557** | 8,59 | y | 154Eu | 4420 | 154Gd | **1028** | -2,149 | 8,8E+07 | 7,95 |
| 62. | 158Gd | -1,086 | **323,6** | 18,479 | h | 159Gd | *455,2* | 159Tb | **1817** | -1,22 | -6,3E+10 | n/a |

| | $nuc_1$ | $\log y_1$ (Si=6) | $\sigma_1$ (mb) | $nuc_2$ | $T_2$ | time unit | $\sigma_2$ (mb) | $nuc_3$ | $\sigma_3$ (mb) | $\log y_3$ (Si=6) | $n_n$ (cm⁻³) | lg $n_n$ |
|---|---|---|---|---|---|---|---|---|---|---|---|---|
| 63. | 159Tb | -1,22 | **1817** | 72,3 | | d | 160Tb | 3240 | 160Dy | **890** | -2,044 | 3,6E+09 | 9,55 |
| 64. | 164Dy | -0,955 | **212** | 2,334 | | h | 165Dy | *284,5* | 165Ho | **1237** | -1,051 | -8,4E+11 | n/a |
| 65. | 165Ho | -1,051 | **1237** | 26,83 | | h | 166Ho | *1262,0* | 166Er | **700** | -1,077 | 2,1E+09 | 9,32 |
| 66. | 165Ho | -1,051 | **1237** | 1,20E+03 | | y | 166Hoᵐ | *1235,3* | 166Er | **700** | -1,077 | 5,4E+03 | 3,73 |
| 67. | 168Er | -1,168 | **319** | 9,392 | | d | 169Er | 653 | 169Tm | **1065** | -1,423 | -2,5E+09 | n/a |
| 68. | 169Tm | -1,423 | **1065** | 128,6 | | d | 170Tm | 1870 | 170Yb | **768,3** | -2,124 | 7,2E+08 | 8,86 |
| 69. | 174Yb | -1,103 | **150,5** | 4,158 | | d | 175Yb | 558 | 175Lu | **1219** | -1,449 | -1,1E+10 | n/a |
| 70. | 176Lu | -2,959 | **1639** | 6,6475 | | d | 177Lu | *794,9* | 177Hf | **1544** | -1,545 | -5,8E+09 | n/a |
| 71. | 176Lu | -2,959 | **1639** | 160,44 | | d | 177Luᵐ | *348,3* | 177Hf | **1544** | -1,545 | -5,5E+08 | n/a |
| 72. | 176Lu | -2,959 | **1639** | 6 | | m | 177Luᵐ | *348,3* | 177Hf | **1544** | -1,545 | -2,1E+13 | n/a |
| 73. | 180Hf | -1,265 | **156,6** | 42,39 | | d | 181Hf | 194 | 181Ta | **766** | -1,684 | -2,2E+09 | n/a |
| 74. | 181Ta | -1,684 | **766** | 114,43 | | d | 182Ta | 1120 | 182W | **285** | -1,455 | 3,0E+06 | 6,48 |
| 75. | 184W | -1,389 | **225** | 75,4 | | d | 185W | 633 | 185Re | **1438,5** | -1,717 | -4,5E+08 | n/a |
| 76. | 186W | -1,423 | **226** | 23,72 | | h | 187W | <u>183</u> | 187Re | **1184** | -1,475 | -1,3E+11 | n/a |
| 77. | 185Re | -1,717 | **1420** | 3,7186 | | d | 186Re | 743 | 186Os | **414** | -2,06 | 7,0E+10 | 10,85 |
| 78. | 185Re | -1,717 | **18,5** | 2,00E+05 | | y* | 186Reᵐ | 743 | 186Os | **414** | -2,06 | -4,9E+02 | n/a |
| 79. | 187Re | -1,475 | **1187** | 17,003 | | h | 188Re | <u>2343</u> | 188Os | **294** | -1,047 | 9,0E+09 | 9,96 |
| 80. | 190Os | -0,75 | **278** | 15,4 | | d | 191Os | 1290 | 191Ir | **1350** | -0,607 | -1,4E+09 | n/a |
| 81. | 192Os | -0,558 | **160** | 30,11 | | h | 193Os | *95,6* | 193Ir | **994** | -0,383 | -2,2E+11 | n/a |
| 82. | 191Ir | -0,607 | **1350** | 7,38E+01 | | d | 192Ir | 2080 | 192Pt | **483** | -1,979 | 1,3E+10 | 10,10 |
| 83. | 191Ir | -0,607 | **1350** | 2,41E+02 | | y* | 192Irᵐ | 2080 | 192Pt | **483** | -1,979 | 1,1E+07 | 7,02 |
| 84. | 193Ir | -0,383 | **994** | 19,28 | | h | 194Ir | *397,8* | 194Pt | **283** | -0,356 | 2,2E+11 | 11,34 |
| 85. | 196Pt | -0,47 | **167,4** | 19,8915 | | h | 197Pt | *175,4* | 197Au | **612,8** | -0,728 | -1,0E+11 | n/a |
| 86. | 197Au | -0,728 | **612,8** | 2,70E+00 | | d | 198Au | 840 | 198Hg | **173** | -1,46 | 2,5E+11 | 11,39 |
| 87. | 202Hg | -0,996 | **63,3** | 46,594 | | d | 203Hg | 98 | 203Tl | **170,5** | -1,265 | -9,8E+10 | n/a |
| 88. | 204Hg | -1,633 | **42** | 5,14 | | m | 205Hg | <u>12,7</u> | 205Tl | **52,6** | -0,886 | -5,6E+14 | n/a |
| 89. | 203Tl | -1,265 | **170,5** | 3,78 | | y | 204Tl | 215 | 204Pb | **83,7** | -1,207 | 7,8E+07 | 7,89 |
| 90. | 205Tl | -0,866 | **52,6** | 4,2 | | m | 206Tl | <u>34,1</u> | 206Pb | **14,7** | -0,277 | -2,3E+13 | n/a |
| 91. | 208Pb | 0,265 | **0,376** | 3,253 | | h | 209Pb | *3,6* | 209Bi | **2,61** | -0,842 | 5,1E+13 | 13,71 |

## 4. Isotonic equilibrium neutron density at a temperature of $3 \cdot 10^8$ K

| | $nuc_1$ | $\log y_1$ (Si=6) | $\sigma_1$ (mb | $T_2$ | time unit | $nuc_2$ | $\sigma_2$ (mb) | $nuc_3$ | $\sigma_3$ (mb) | $\log y_3$ (Si=6) | $n_n$ (cm⁻³) | lg $n_n$ |
|---|---|---|---|---|---|---|---|---|---|---|---|---|
| 1. | 58Fe | 3,473 | **13,3** | 44,4495 | d | 59Fe | **26,4** | 59Co | **40,1** | 3,352 | -1,41E+10 | n/a |
| 2. | 59Co | 3,352 | **40,1** | 1925,28 | d | 60Co | *12,64* | 60Ni | **29,9** | 4,11 | -2,85E+10 | n/a |
| 3. | 62Ni | 3,255 | **22,2** | 100,1 | y | 63Ni | **66,7** | 63Cu | **53,7** | 2,558 | 9,50E+07 | 7,98 |

| | nuc$_1$ | log y$_1$ (Si=6) | σ$_1$(mb | T$_2$ | time unit | nuc$_2$ | σ$_2$(mb) | nuc$_3$ | σ$_3$(mb) | log y$_3$ (Si=6) | n$_n$ (cm$^{-3}$) | lg n$_n$ |
|---|---|---|---|---|---|---|---|---|---|---|---|---|
| 4. | 64Ni | 2,726 | 8 | 2,5172 | h | 65Ni | 21,2 | 65Cu | 29 | 2,207 | -1,18E+12 | n/a |
| 5. | 65Cu | 2,207 | 29 | 5,12 | m | 66Cu | 126,6 | 66Zn | 36,4 | 2,544 | -4,14E+13 | n/a |
| 6. | 68Zn | 2,369 | 20,7 | 56,4 | m | 69Zn | 75,4 | 69Ga | 118,7 | 1,358 | 7,89E+12 | 12,90 |
| 7. | 68Zn | 2,369 | 20,7 | 13,76 | h* | 69Zn$^m$ | 110,9 | 69Ga | 118,7 | 1,358 | 3,67E+11 | 11,56 |
| 8. | 70Zn | 0,893 | 10,9 | 2,45 | m | 71Zn | 48,26 | 71Ga | 106,4 | 1,176 | -3,37E+14 | n/a |
| 9. | 70Zn | 0,893 | 10,9 | 3,96 | h | 71Zn$^m$ | 48,26 | 71Ga | 106,4 | 1,176 | -3,48E+12 | n/a |
| 10. | 69Ga | 1,358 | 118,7 | 21,14 | m | 70Ga | 301,5 | 70Ge | 89,1 | 1,387 | 1,64E+12 | 12,22 |
| 11. | 71Ga | 1,176 | 106,4 | 14,095 | h | 72Ga | 267,6 | 72Ge | 53,1 | 1,513 | -7,81E+10 | n/a |
| 12. | 74Ge | 1,638 | 37,4 | 82,78 | m | 75Ge | 203,1 | 75As | 355 | 0,817 | -7,49E+11 | n/a |
| 13. | 75As | 0,817 | 355 | 1,0942 | d | 76As | 469,6 | 76Se | 168 | 0,748 | 1,40E+13 | 13,15 |
| 14. | 78Se | 1,164 | 61,1 | 2,95E+05 | y | 79Se | 263 | 79Br | 622 | 0,775 | -1,39E+07 | n/a |
| 15. | 80Se | 1,49 | 38 | 18,45 | m | 81Se | 229 | 81Br | 239 | 0,766 | -1,48E+12 | n/a |
| 16. | 80Se | 1,49 | 38 | 57,28 | m* | 81Se$^m$ | 229 | 81Br | 239 | 0,766 | -4,78E+11 | n/a |
| 17. | 79Br | 0,775 | 622 | 17,68 | m | 80Br | 790 | 80Kr | 274 | 0,009 | 2,42E+13 | 13,38 |
| 18. | 79Br | 0,775 | 622 | 4,4205 | h* | 80Br$^m$ | 790 | 80Kr | 90,4 | 0,009 | 5,15E+12 | 12,71 |
| 19. | 81Br | 0,766 | 239 | 35,282 | h | 82Br | 390,5 | 82Kr | 93 | 0,716 | 1,05E+12 | 12,02 |
| 20. | 84Kr | 1,408 | 32,6 | 3916,8 | d | 85Kr | 73 | 85Rb | 234 | 0,709 | -3,23E+07 | n/a |
| 21. | 84Kr | 1,408 | 32,6 | 4,78 | h79% | 85Kr$^m$ | 73 | 85Rb | 234 | 0,709 | -6,35E+11 | n/a |
| 22. | 86Kr | 0,893 | 4,76 | 76,3 | m | 87Kr | 18 | 87Rb | 15,7 | 0,316 | 4,48E+12 | 12,65 |
| 23. | 85Rb | 0,709 | 234 | 18,642 | d | 86Rb | 202 | 86Sr | 63,5 | 0,365 | 5,60E+10 | 10,75 |
| 24. | 87Rb | 0,316 | 15,7 | 17,773 | m | 88Rb | 110 | 88Sr | 6,16 | 1,288 | -1,14E+13 | n/a |
| 25. | 88Sr | 1,288 | 6,16 | 50,57 | d | 89Sr | 19 | 89Y | 19,3 | 0,667 | 1,03E+10 | 10,01 |
| 26. | 89Y | 0,667 | 19,3 | 90 | y | 90Y | 149,9 | 90Zr | 19,3 | 0,769 | -1,26E+06 | n/a |
| 27. | 92Zr | 0,299 | 37,8 | 1,53E+06 | y | 93Zr | 96 | 93Nb | 265,7 | 0,299 | -1,58E+03 | n/a |
| 28. | 93Nb | -0,156 | 265,7 | 2,03E+04 | y | 94Nb | 482 | 94Mo | 109,6 | -0,636 | 4,04E+10 | 10,61 |
| 29. | 102Ru | -0,231 | 151 | 39,26 | d | 103Ru | 343 | 103Rh | 810 | -0,463 | -3,59E+10 | n/a |
| 30. | 103Rh | -0,463 | 810 | 42,3 | s | 104Rh | 154 | 104Pd | 274 | -0,818 | 1,73E+15 | 15,24 |
| 31. | 103Rh | -0,463 | 810 | 4,34 | m* | 104Rh$^m$ | 154 | 104Pd | 274 | -0,818 | 2,80E+14 | 14,45 |
| 32. | 106Pd | -0,42 | 244 | 6,60E+06 | y | 107Pd | 1302 | 107Ag | 787 | -0,602 | -4,02E+04 | n/a |
| 33. | 108Pd | -0,431 | 218 | 13,7012 | h | 109Pd | 236,3 | 109Ag | 793 | -0,627 | -1,24E+11 | n/a |
| 34. | 107Ag | -0,602 | 787 | 2,37 | m | 108Ag | 1788 | 108Cd | 202 | -1,851 | 5,96E+14 | 14,77 |
| 35. | 107Ag | -0,602 | 787 | 438 | y* | 108Ag$^m$ | 1383,0 | 108Cd | 202 | -1,851 | 7,92E+06 | 6,90 |
| 36. | 109Ag | -0,627 | 793 | 24,6 | s | 110Ag | 1172 | 110Cd | 229,9 | -0,701 | 1,03E+14 | 14,01 |
| 37. | 109Ag | -0,627 | 793 | 249,76 | d | 110Ag$^m$ | 1172 | 110Cd | 229,9 | -0,701 | 1,18E+08 | 8,07 |
| 38. | 114Cd | -0,333 | 135,3 | 53,46 | h | 115Cd | 290 | 115In | 776 | -0,754 | -2,47E+10 | n/a |

| | nuc$_1$ | log y$_1$ (Si=6) | σ$_1$ (mb) | T$_2$ | time unit | nuc$_2$ | σ$_2$ (mb) | nuc$_3$ | σ$_3$ (mb) | log y$_3$ (Si=6) | n$_n$ (cm$^{-3}$) | lg n$_n$ |
|---|---|---|---|---|---|---|---|---|---|---|---|---|
| 39. | 114Cd | -0,333 | **135,3** | 44,56 | d* | 115Cd$^m$ | 224 | 115In | 776 | -0,754 | -1,60E+09 | n/a |
| 40. | 113In | -2,103 | **229** | 71,9 | s | 114In | <u>1308</u> | 114Sn | **134,4** | -1,599 | -1,26E+13 | n/a |
| 41. | 113In | -2,103 | **580** | 49,51 | d* | 114In$^m$ | 2595 | 114Sn | **134,4** | -1,599 | 8,07E+07 | 7,91 |
| 42. | 115In | -0,754 | **154** | 14,1 | s | 116In | <u>1377</u> | 116Sn | **92,3** | -0,263 | -5,91E+13 | n/a |
| 43. | 115In | -0,754 | **622** | 54,29 | m | 116In$^m$ | <u>1377</u> | 116Sn | **92,3** | -0,263 | 6,51E+11 | 11,81 |
| 44. | 120Sn | 0,097 | **36,3** | 27,03 | h | 121Sn | 167 | 121Sb | **532** | -0,752 | -2,40E+10 | n/a |
| 45. | 120Sn | 0,097 | **36,3** | 43,9 | y* | 121Sn$^m$ | *175,4* | 121Sb | **532** | -0,752 | -1,60E+06 | n/a |
| 46. | 122Sn | -0,745 | **22,6** | 129,2 | d | 123Sn | 361 | 123Sb | **303** | -0,879 | -3,00E+11 | n/a |
| 47. | 122Sn | -0,745 | **21,7** | 40,06 | m | 123Sn$^m$ | 361 | 123Sb | **303** | -0,879 | -1,40E+15 | n/a |
| 48. | 121Sb | -0,752 | **532** | 2,7238 | d | 122Sb | 894 | 122Te | **295,4** | -0,928 | 7,76E+10 | 10,89 |
| 49. | 123Sb | -0,879 | **303** | 60,11 | d | 124Sb | *924,0* | 124Te | **155** | -0,654 | 5,00E+11 | 11,70 |
| 50. | 126Te | -0,046 | **81,3** | 9,35 | h | 127Te | *256,8* | 127I | **662** | -0,046 | -2,23E+11 | n/a |
| 51. | 126Te | -0,046 | **81,3** | 109 | d* | 127Te$^m$ | *668,6* | 127I | **662** | -0,046 | -3,06E+08 | n/a |
| 52. | 127I | -0,046 | **662** | 24,99 | m | 128I | *679,5* | 128Xe | **262,5** | 0,991 | -1,83E+12 | n/a |
| 53. | 132Xe | 0,086 | **63,8** | 5,243 | d | 133Xe | 127 | 133Cs | **502** | -0,429 | -3,64E+10 | n/a |
| 54. | 133Cs | -0,429 | **502** | 2,0652 | y | 134Cs | 724 | 134Ba | **176** | -0,963 | 9,52E+10 | 10,98 |
| 55. | 138Ba | 0,508 | **4,13** | 83,06 | m | 139Ba | *39,7* | 139La | **32,4** | -0,351 | -1,02E+12 | n/a |
| 56. | 139La | -0,351 | **32,4** | 1,67855 | d | 140La | 117,8 | 140Ce | **11,73** | 0,004 | 4,11E+10 | 10,61 |
| 57. | 140Ce | 0,004 | **11,73** | 32,508 | d | 141Ce | 76 | 141Pr | **111,4** | -0,777 | -4,35E+09 | n/a |
| 58. | 141Pr | -0,777 | **111,4** | 19,12 | h | 142Pr | 361,4 | 142Nd | **35,1** | -0,65 | 3,60E+10 | 10,56 |
| 59. | 150Sm | -1,717 | **422,3** | 90 | y | 151Sm | 3040 | 151Eu | **3556** | 0,35 | -7,31E+06 | n/a |
| 60. | 152Sm | -1,161 | **464,8** | 46,284 | h | 153Sm | 1095 | 153Eu | **2567** | -1,294 | -1,81E+10 | n/a |
| 61. | 153Eu | -1,294 | **2557** | 8,59 | y | 154Eu | 4420 | 154Gd | **1028** | -2,149 | 1,08E+10 | 10,03 |
| 62. | 158Gd | -1,086 | **323,6** | 18,479 | h | 159Gd | *455,2* | 159Tb | **1817** | -1,22 | -6,09E+10 | n/a |
| 63. | 159Tb | -1,22 | **1817** | 72,3 | d | 160Tb | 3240 | 160Dy | **890** | -2,044 | 2,52E+11 | 11,40 |
| 64. | 164Dy | -0,955 | **212** | 2,334 | h | 165Dy | *284,5* | 165Ho | **1237** | -1,051 | -8,43E+11 | n/a |
| 65. | 165Ho | -1,051 | **1237** | 26,83 | h | 166Ho | *1262,0* | 166Er | **700** | -1,077 | 1,16E+10 | 10,07 |
| 66. | 165Ho | -1,051 | **1237** | 1,20E+03 | y | 166Ho$^m$ | *1235,3* | 166Er | **700** | -1,077 | 3,03E+04 | 4,48 |
| 67. | 168Er | -1,168 | **319** | 9,392 | d | 169Er | 653 | 169Tm | **1065** | -1,423 | -2,40E+09 | n/a |
| 68. | 169Tm | -1,423 | **1065** | 128,6 | d | 170Tm | 1870 | 170Yb | **768,3** | -2,124 | 7,95E+08 | 8,90 |
| 69. | 174Yb | -1,103 | **150,5** | 4,158 | d | 175Yb | 558 | 175Lu | **1219** | -1,449 | -1,64E+10 | n/a |
| 70. | 176Lu | -2,959 | **1639** | 6,6475 | d | 177Lu | *794,9* | 177Hf | **1544** | -1,545 | -8,18E+09 | n/a |
| 71. | 176Lu | -2,959 | **1639** | 160,44 | d | 177Lu$^m$ | *348,3* | 177Hf | **1544** | -1,545 | -7,73E+08 | n/a |
| 72. | 176Lu | -2,959 | **1639** | 6 | m | 177Lu$^m$ | *348,3* | 177Hf | **1544** | -1,545 | -2,98E+13 | n/a |
| 73. | 180Hf | -1,265 | **156,6** | 42,39 | d | 181Hf | 194 | 181Ta | **766** | -1,684 | -5,63E+10 | n/a |
| 74. | 181Ta | -1,684 | **766** | 114,43 | d | 182Ta | 1120 | 182W | **285** | -1,455 | 5,21E+08 | 8,72 |
| 75. | 184W | -1,389 | **225** | 75,4 | d | 185W | 633 | 185Re | **1438,5** | -1,717 | -4,83E+08 | n/a |

| | nuc$_1$ | log y$_1$ (Si=6) | σ$_1$(mb | T$_2$ | time unit | nuc$_2$ | σ$_2$(mb) | nuc$_3$ | σ$_3$(mb) | log y$_3$ (Si=6) | n$_n$ (cm$^{-3}$) | lg n$_n$ |
|---|---|---|---|---|---|---|---|---|---|---|---|---|
| 76. | 186W | -1,423 | **226** | 23,72 | h | 187W | <u>183</u> | 187Re | **1184** | -1,475 | -1,41E+11 | n/a |
| 77. | 185Re | -1,717 | **1420** | 3,7186 | d | 186Re | 743 | 186Os | **414** | -2,06 | 7,48E+10 | 10,87 |
| 78. | 185Re | -1,717 | **18,5** | 2,00E+05 | y* | 186Re$^m$ | 743 | 186Os | **414** | -2,06 | -5,23E+02 | n/a |
| 79. | 187Re | -1,475 | **1187** | 17,003 | h | 188Re | <u>2343</u> | 188Os | **294** | -1,047 | 9,50E+09 | 9,98 |
| 80. | 190Os | -0,75 | **278** | 15,4 | d | 191Os | 1290 | 191Ir | **1350** | -0,607 | -2,58E+09 | n/a |
| 81. | 192Os | -0,558 | **160** | 30,11 | h | 193Os | *95,6* | 193Ir | **994** | -0,383 | -2,05E+11 | n/a |
| 82. | 191Ir | -0,607 | **1350** | 7,38E+01 | d | 192Ir | 2080 | 192Pt | **483** | -1,979 | 1,76E+10 | 10,24 |
| 83. | 191Ir | -0,607 | **1350** | 2,41E+02 | y* | 192Ir$^m$ | 2080 | 192Pt | **483** | -1,979 | 1,47E+07 | 7,17 |
| 84. | 193Ir | -0,383 | **994** | 19,28 | h | 194Ir | *397,8* | 194Pt | **283** | -0,356 | 3,47E+11 | 11,54 |
| 85. | 196Pt | -0,47 | **167,4** | 19,8915 | h | 197Pt | *175,4* | 197Au | **612,8** | -0,728 | -1,19E+11 | n/a |
| 86. | 197Au | -0,728 | **612,8** | 2,70E+00 | d | 198Au | 840 | 198Hg | **173** | -1,46 | 9,52E+11 | 11,98 |
| 87. | 202Hg | -0,996 | **63,3** | 46,594 | d | 203Hg | 98 | 203Tl | **170,5** | -1,265 | -1,40E+11 | n/a |
| 88. | 204Hg | -1,633 | **42** | 5,14 | m | 205Hg | <u>12,7</u> | 205Tl | **52,6** | -0,886 | -5,59E+14 | n/a |
| 89. | 203Tl | -1,265 | **170,5** | 3,78 | y | 204Tl | 215 | 204Pb | **83,7** | -1,207 | 1,55E+10 | 10,19 |
| 90. | 205Tl | -0,866 | **52,6** | 4,2 | m | 206Tl | 34,1 | 206Pb | **14,7** | -0,277 | -2,32E+13 | n/a |
| 91. | 208Pb | 0,265 | **0,376** | 3,253 | h | 209Pb | *3,6* | 209Bi | **2,61** | -0,842 | 5,10E+13 | 13,71 |

## Legend

n/a:  not all third nuclei can be achieved in that process

log y$_i$:  D. Arnett: *Supernovae and Nucleosynthesis*, Princeton University Press, 1996

at σ$_i$ :

**bold**:  KADoNiS 1.0

normal:  KADoNiS 0.3 (MACS)

*italic*:  B. Pritychenko,∗ S.F. Mughabghab: Neutron Thermal Cross Sections, Westcott Factors, Resonance Integrals, Maxwellian Averaged Cross Sections and Astrophysical Reaction Rates Calculated from the ENDF/B-VII.1, JEFF-3.1.2, JENDL-4.0, ROSFOND-2010, CENDL-3.1 and EAF-2010 Evaluated Data Libraries

<u>underlined</u>:  http://adg.llnl.gov/Research/RRSN/aemr/30kev/rath00_7.4.30kev_calc